\documentclass[prc,unsortedaddress,superscriptaddress,twocolumn,amssymb,showpacs,floatfix,nofootinbib]{revtex4-1}
\usepackage{graphics,epsfig}
\usepackage{amsmath}
\usepackage{dcolumn}   

\begin{document}
\date{\today}
\title{Resonance properties  including asymptotic normalization coefficients deduced from phase-shift data without the effective-range function}

\author{B. F. Irgaziev}\email{irgaziev@yahoo.com}
\affiliation{National University of Uzbekistan, Tashkent,Uzbekistan}
\affiliation{GIK Institute of Engineering Sciences and Technology, Topi, Pakistan}
\author{Yu. V. Orlov}\email{orlov@srd.sinp.msu.ru}
\affiliation{Skobeltsyn Nuclear Physics Institute, Lomonosov Moscow State
University, Russia}
\begin{abstract}
 Recently, a new $\Delta$ method for the calculation of asymptotic normalization coefficients (ANC) from phase-shift data has been formulated, proved and used for bound states. This method differs from the conventional one by fitting only the nuclear part of the effective-range function which includes a partial phase shift. It should be applied to large-charge nuclei when the conventional effective-range expansion or the Pad\'e-approximations using the effective-range function $K_l(k^2)$ fitting do not work.  A typical example is the nucleus vertex $\alpha+^{12}$C $\longleftrightarrow ^{16}$O.  Here we apply the $\Delta$ method, which totally excludes the effective-range function, to isolated resonance states. In fact, we return to the initial renormalized scattering amplitude with a denominator which defines the well-known pole condition. Concrete calculations are made for the resonances observed in the  $^3$He-$^4$He, $\alpha$-$\alpha$, and $\alpha$-$^{12}$C collisions. We use the experimental phase-shift and resonant energy data including their uncertainties and find the ANC variations for the states considered.
The corresponding results are in a good agreement with those for the $S$-matrix pole  method which uses the differing formalism. The simple formula for narrow resonances given in the literature is used to check the deduced results. The related ANC function clearly depends on the resonance energy ($E_0$) and width ($\Gamma$), which is used to find the ANC uncertainty ($\Delta$ANC) through  the energy ($\Delta E_0$) and the width ($\Delta\Gamma$) uncertainties.We also discuss the $\Delta$ method differences between bound and resonance states pole conditions.
\end{abstract}

\maketitle

\section{Introduction}\label{intr}
It is known that many reactions in supernovae explosions proceed through sub-threshold bound states and low-energy resonance states. To calculate the rate of such reactions, one needs to find the asymptotic normalization coefficient (ANC) of the radial wave function for bound and resonance states, which can be used to calculate radiative capture cross sections at low energies. The radiative capture process is one of the main sources of new element creation.

In our recent paper \cite{OrlovIrgazievPRC96}, we  validate a new algorithm for the bound states ANC calculation when the input data include a phase-shift energy dependence at low-energy region and bound-state pole position. The related form of the renormalized scattering amplitude is proposed (without proof) in Ref. \cite{spren16} (see also Ref. \cite{blokh16}). We call this algorithm the $\Delta$ method. This new method allows us to avoid the problems arising when the charges of colliding particles increase.  In Ref. \cite{OrlovIrgazievPRC96}, we note that the effective-range expansion (ERE) and Pad\'e approximations for finding the ANC are especially limited by the values of the colliding particle charges. These approaches do not work for large charges when the nuclear term of the effective-range function (ERF) is too small compared with the Coulomb term. The $\alpha ^{12}$C is just a proper example of  such a situation. This problem is revealed in our work \cite{orlov16} where we see that the  ERF for $\alpha ^{12}$C have a similar  behavior in the different $^{16}$O states because the Coulomb term in the ERF does not depend on nuclear state.

In the present paper, we apply the $\Delta$ method to resonance states. The first attempt to use the ERF approach for resonances is made in \cite{ENO}. In \cite{MIGOQ} the problem of calculating  resonance pole properties is solved using the $S$-matrix pole  approach in the frame of a potential model.

In the $S$-matrix pole method (SMP) (see Ref. \cite{irgaz-14}), an analytical continuation to the resonance pole is accomplished for the so-called 'potential' (or no resonant) phase shift in the complex $k$ plane.
 Having found the ANC, we now know   the asymptotic part of
the Gamow wave function. This allows us to normalize it correctly if
we choose a nuclear potential of the interaction between the
two nuclei considered. In Refs. \cite{MIGOQ, irgaz-14}  there is a more detailed discussion about the Gamow states and its normalization.

The ANC method has been explored as an indirect experimental method for determining the cross sections of peripheral reactions at low energy \cite{muk01}. There are several methods of deriving the bound state ANCs from experimental data (see Refs.\cite{muk07,akram99} and references therein). Recently, the  ERE-method has been developed to find the ANC for bound states from the elastic scattering phase-shift $\delta_l$ analysis (see Refs.\cite{OrlIrgNik, spren} and references therein). A renormalized scattering amplitude, taking into account the Coulomb interaction, is derived in Ref. \cite{Hamilton} to enable an analytic continuation of this amplitude to negative energies. It is shown in Ref. \cite{Hamilton} that the ERF is a real analytic function with the possible exception of single poles. This means that the ERF can be presented by the ERE or Pad\'e-approximations, whose coefficients can be found by fitting the experimental phase shifts. The same is valid for the $\Delta$ method as long as the  $\Delta_l$ function is included in one of the two terms defining the ERF.

 An important step in calculating a bound state ANC using the ERF  was first taken by Iwinski \textit{et al.} \cite{Iwinsky}, who discuss a radiative capture process $^3$He($^4$He, $\gamma$)$^7$Be and calculate ANCs for the bound $P$ states with  total angular momentum $J$ and  parity $\pi$ ($J^\pi$=3/2$^-$ and 1/2$^-$)  of the  $^7$Be nucleus applying the Pad\'e-approximant.
 In the past, the results of the scattering phase-shifts analysis were often presented in the form of the ERF. Consequently from 1984 onwards, the ERF has been used to deduce the ANC. It is much simpler to use the few ERF parameters instead of the $\delta_l$ tables. A first example is the nucleon-nucleon scattering when the expansion coefficients are considered as independent characteristics of the $NN$ interaction. The scattering length $a_l$, effective range $r_l$, and the shape parameter $P_l$ for the orbital momentum $l$ were introduced, although ERE should be convergent, and consequently may include an unlimited number of terms. The denomination of the coefficients in the polynomial $K_l(k^2)=-1/a_l + (r_l/2)k^2-P_l r_l^3 k^4$ is due to the fact that one can truncate the series in the low-energy region. Another approximation for $K_l(k^2)$ with a limited number of fitting parameters is the Pad\'e-approximant when the ERF has poles.

In fact, it is necessary to fit only the nuclear part. Excluding the ERF leads to the original renormalized scattering amplitude form which does not include the Coulomb part of the ERF. Simple algebra gives an inverse transformation from the amplitude including the ERF to this original form. When charges are large enough, the original renormalized scattering amplitude form should be used to deduce resonance properties, including ANC, from the experimental phase shifts. This can be used for smaller charges as well.

Below we apply the $\Delta$ method to the concrete systems $^3$He$\alpha$, $\alpha\alpha$, and $\alpha ^{12}$C.
Processes such as the scattering of $\alpha$ particles, triple-$\alpha$ reaction, and radiative $\alpha$ capture play a major role in stellar nucleosynthesis. In particular, $\alpha$  capture on carbon determines the ratio of carbon to oxygen during helium burning, and affects subsequent carbon, neon, oxygen, and silicon burning stages.
The authors of a recent paper \cite{Nature} describe an \textit{ab initio} calculation of $\alpha$-$\alpha$ scattering that uses lattice Monte Carlo simulations to a two-cluster system.

The article is organized as follows. In Sec. II we present the main formulas of the $\Delta$ method for resonances and show that the original renormalized scattering amplitude should be used, which can be analytically continued to a resonant pole. It is important that this amplitude does not include the ERF and its Coulomb part. The ratio of  Coulomb to nuclear parts increases quickly with the growth of the product of the colliding nuclei charges. In our paper \cite{OrlovIrgazievPRC96}, one can see from Eqs. (9) and (12) that the nuclear term includes exp(2$\pi\eta$) in the denominator. That is why the relatively small variation of $\eta$ (or charges product) leads to a strong reduction of the nuclear component compared with the Coulomb term in square brackets in Eq. (9) in Ref. \cite{OrlovIrgazievPRC96}. It is notable that $\eta$ = 1/$(a_B k)$ is the only argument of the functions responsible for the Coulomb effects. Here $a_B$ is the Bohr radius. $ \eta$ has a scaling property: a decrease of  $a_B$ is equivalent to the same decrease of the relative momentum $k$ of  the colliding nuclei and the corresponding decrease of the energy when the role of the Coulomb barrier increases.
 As a result, the nuclear part, including the phase shift, is a small addition to the Coulomb part $h(\eta)$ which can be ignored with  reasonable precision. The ratio of the Coulomb/nuclear parts is about 10$^3$ for the $\alpha$-$^{12}$C  system due to a relatively large value of the Sommerfeld parameter $\eta$.
As a result, the corresponding phase shift is unreproducible from the experimental ERF fit which leads to an incorrectly calculated ANC. By definition, the $\Delta_l$ function fit reproduces the input phase shift. That is why we named the corresponding algorithm the $\Delta$ method. So the final equation for the renormalized scattering amplitude can be applied to calculate the nuclear vertex constant (NVC) or $\tilde{G}_l$, the residue $W_l$ and the ANC ($C_l$).
The relationships between these observables are well known in the literature. The simple analytic ANC formula for narrow resonances is written, borrowed from Ref. \cite {dolin_akram}. This simple formula clearly depends on the resonance energy and width. The ANC uncertainty equation is due to the uncertainties in the $E_0$ and $\Gamma$ and is derived from this ANC expression. We also clarify some points concerning the $\Delta$-method validation for a bound state considered in  Ref. \cite{OrlovIrgazievPRC96} and explain why the bound state pole condition differs from that for a resonance.

In Sec. III we present the main SMP-method equations which describe a different formalism compared with the $\Delta$ method. The only common elements in both approaches are the one-channel approximation and  model which does not take into account the internal structure of colliding nuclei. The SMP-method results are published in  Ref. \cite{irgaz-14} for the resonance states of $^5$He, $^5$Li, and $^{16}$O.

In Sec. IV the results are given for the $\Delta$ method calculations for the resonance levels of $^7$Be, $^8$Be, and $^{16}$O. Tables \ref{tab1}--\ref{tab3} for the three nuclear systems studied here include the experimental and calculated resonant energies $E_0$ and the widths $\Gamma$ for the different methods. The resulting ANCs  are compared with those calculated by the SMP-method and with those  using the simple  formula for narrow resonances.
 These tables show a good agreement between the results obtained by both of these methods. The results for narrow resonances  serve to check our $\Delta$ method calculation results. We give the absolute values $\mid C_l\mid$ because the Schr$\rm{\ddot{o}}$dinger equation is uniform, so the phase multiplier can be omitted.

The effects in the calculated ANCs of the experimental uncertainties in the phase shift and resonant energy are investigated.
The conclusions following from the analysis of the results of all the tables are formulated. We stress the stability of the different results found for the $^8$Be ground state, which plays a special role in astrophysics.

In Sec. V  we summarize the main points of the present paper

In the following we use the unit system $\hbar=c=1$.

\section{$\Delta$ method for resonant states without using the effective-range function}\label{no_eff_range}

The partial amplitude of the nuclear scattering in the presence of the Coulomb interaction is
\begin{equation}\label{amp}
f_l(k)=\exp(2i\sigma_l) [\exp(2i\delta_l)-1]/2ik,
\end{equation}
where
 \begin{equation} \label{Coulomb phase}
 \exp(2i\sigma_l) =
 {\Gamma(l+1+i\eta)}/{\Gamma(l+1-i\eta)},
\end{equation}

Here $\delta_l$ is the nuclear phase shift modified by the Coulomb interaction, and $\eta=\xi/k$ is the Sommerfeld parameter, $\xi=Z_1Z_2\mu\alpha$, $k=\sqrt{2\mu E}$ is the relative momentum; $\mu$ and $E$  are the reduced mass and the center-of-mass  energy  of the colliding nuclei with the charge numbers $Z_1$ and $Z_2$,  respectively; and $\alpha$ is the fine-structure constant.

The amplitude (\ref{amp}) has a complicated analytical property in the complex momentum plane due to the Coulomb factor. According to Ref. \cite{Hamilton}, we renormalize  the partial amplitude of the elastic scattering multiplying it by the function
[the Coulomb correcting or re-normalizing factor $CF_l(k)$]
\begin{equation} \label{yost}
CF_l(k)=\frac{(l!)^2e^{\pi\eta}}{(\Gamma(l+1+i\eta))^2}.
\end{equation}

The general pole condition $\cot\delta_l-i=0$ follows from the expression for the renormalized amplitude of the elastic scattering (see, for example, Ref. \cite{OrlovIrgazievPRC96})
\begin{equation}\label{fl2}
\tilde{f}_l(k)=\frac{1}{k(\cot\delta_l-i)\rho_l(k)},
\end{equation}
where the function $\rho_l$ is defined by the equation
\begin{equation}\label{rho}
\rho_l(k)=\frac{2\pi\eta}{e^{2\pi\eta}-1}\prod_{n=1}^l\Bigl(1+\frac{\eta^2}{n^2}\Bigr).
\end{equation}

Writing the expression $\cot\delta_l$ in a nonphysical energy region in  Eq.(\ref{fl2}) and elsewhere, we mean its analytical continuation, since the phase shift is defined only in the positive energy region.
The renormalized scattering amplitude of the conventional method is written as
\begin{equation}\label{renormalized amplitude}
 \tilde f_l (k) = \frac{k^{2l}}{K_l(k^2)-2\xi D_l(k^2)h(\eta)}
\end{equation}
(see, for example, Ref. \cite{OrlIrgNik} and definitions below), where the effective-range function $K_l(k^2)$
 borrowed from Ref. \cite{haer} has the form:
\begin{equation} \label{CoulombKl}
K_l(k^2) = 2 \xi D_l(k^2)\left[C_0^2(\eta)(\cot\delta_l - i)+h(\eta )\right],
\end{equation}
\begin{equation} \label{Coulomb_h}
  h(\eta)\equiv \Psi (i\eta ) + (2i\eta )^{- 1} - \ln(i\eta),
\end{equation}
where $\Psi (i\eta )$ is the digamma function.

It is easy to derive (\ref{fl2}), substituting (\ref{CoulombKl}) into the denominator of (\ref{renormalized amplitude}). Using simple algebra we obtain the expression
\begin{equation}\label{fl3}
\tilde{f}_l(k)=\frac{k^{2l}}{2\xi D_l(k^2)C_0^2(\eta)(\cot\delta_l-i)},
\end{equation}
where the function $h(\eta )$ (\ref{Coulomb_h}) is absent.
In Eqs. (\ref{renormalized amplitude})--(\ref{fl3})
we use the following notations:
\begin{eqnarray}\label{CandDfunctions}
C_0^2(\eta)&=&\frac{\pi} {\exp(2\pi\eta)-1},\label{C02}\\
D_l(k^2)&=&\prod_{n=1}^l\Bigl(k^2+\frac{\xi^2}{n^2}\Bigr),\qquad D_0(k^2)=1 \label{DL_k}.
\end{eqnarray}
We define the $\Delta_l(k^2)$ function as in Ref. \cite{OrlovIrgazievPRC96}
\begin{equation}\label{Delta-R}
\Delta_l(k^2)=C_0^2(\eta)\cot\delta_l
\end{equation}
in the positive energy semi-axis.
Using (\ref{Delta-R}) we can recast (\ref{fl3}) as
\begin{equation}\label{fl4}
\tilde{f}_l(k)=\frac{k^{2l}}{2\xi D_l(k^2)[\Delta_l(k^2)-iC_0^2(\eta)]}.
\end{equation}
We note that $C_0^2(\eta)\rightarrow k/2\xi$ and $D_l(k^2)\rightarrow k^{2l}$, if $\eta=\xi/k \rightarrow 0$. Therefore $\tilde f_l (k) \rightarrow 1/k(\rm{cot} \delta_l-i)$ and $K_l(k^2)\rightarrow k^{2l+1}\rm{cot} \delta_l$, as it should be when there is no  Coulomb interaction. The factor $C_0^2(\eta)$ secures a regularity of the  $\Delta _l$-function at point $E$=0. The physical meaning of the function $C_0^2(\eta)$  is its role as the compensating factor, excluding the essential phase-shift singularity in the function $\delta_l$.  Moreover this is a multiplier in the Coulomb penetration factor squared (see Eq. (\ref{rho})). Separating this factor from the total partial amplitude leads to the renormalized amplitude (or ‘effective amplitude’ as it is called in Ref. \cite{Hamilton}). This has analytic properties similar to amplitude properties for a short-range potential.

It is easy to show that the expression (\ref{fl4}) is equivalent to the original  formula (\ref{fl2}), although this is obvious because (\ref{fl4}) is derived from (\ref{fl2}) and from the expression (\ref{CoulombKl}) for $K_l(k^2)$.
To prove this, we express the function $k\rho(k)$ in terms of $C_0^2(\eta)$ and $D_l(k^2)$ as
$$
k\rho(k)=2\xi C_0^2(\eta)D_l(k^2)/k^{2l}
$$
and include it in Eq. (\ref{fl2}).
The function $C_0^2(\eta)$, having the analytical form  (\ref{CandDfunctions}), does not need fitting. This function clearly depends on the momentum $k$ through $\eta(k)$ which leads to the square root cut of the renormalized amplitude in the $E$ plane. 

We would like to clarify here some points concerning the $\Delta$ method validation for bound states discussed in Ref. \cite{OrlovIrgazievPRC96}. The renormalized amplitude $\tilde{f}_l(k)$ (which is called ‘effective’ in Ref. \cite{Hamilton}) is written in terms of the effective-range function $K_l(k^2)$  in Eq. (\ref{renormalized amplitude}). It is noted in Ref. \cite{Hamilton} that this amplitude has the analytical properties similar to those for a short-range potential.  This means that the renormalized amplitude $\tilde{f}_l(k)$  is real in the imaginary positive semi-axis of the complex $k$ plane, i.e. for negative energies $E$. It has a logarithmic cut $-\infty<E\leq E_s $ where $E_s$  is the position of a nearest singularity for an exchange Feynman diagram and the square root cut $0\leq E<\infty$ in the complex $E$ plane. The numerator in Eq.  (\ref{renormalized amplitude}), $k^{2l} \sim E^l$, clearly depends on energy. So, the denominator in  Eq.  (\ref{renormalized amplitude}) is a function of $k$, having the square root cut $0\leq E <\infty$ in the complex $E$ plane. This function is real for negative energies $E_s\leq E\leq 0$ and becomes complex on the square root cut. These properties correspond to the procedure of the continuation   from positive to negative energies where it can be written as
\begin{equation} \label{denom}
 K_l(k^2)-2\xi D_l(k^2)h(\eta)= 2\xi D_l(k^2)\Delta_l(k^2), \, E<0.
\end{equation}
It follows from Eq. (\ref{denom}) that $\Delta_l (k^2)$ is a real analytic function.  Hence, the function $\Delta_l(k^2)$  is real in the interval $E_s\leq E \leq 0$ of negative energy because the functions $h(\eta)$ and $K_l(k^2)$  are real for $E<0$. The denominator in (\ref{renormalized amplitude}) is equal to zero when $\Delta_l (k^2)$=0 or $D_l(k^2)$=0. The last equation gives zeroes at negative energies which do not depend on a nuclear interaction. So we confirm the bound state pole condition  $\Delta_l (k^2)$=0 deduced in Ref. \cite{OrlovIrgazievPRC96}.

The results of the $\Delta_l(k^2)$ fitting using  the experimental phase shifts can be applied to resonances, taking into account that the resonance energy position is defined by the condition
\begin{equation}\label{resonance pole condition}
\Delta_l(k^2)-iC_0^2(\eta)=0.
\end{equation}

 Next we write down the expression for the residue of the $\tilde{f}_l(k)$ at the resonance pole.
This residue of the renormalized amplitude can be written as
\begin{equation}\label{ResidueForResonance}
W_l(k_r^2)=\frac{(k_r)^{2l}}{2\xi D_l(k_r^2)\lim_{k\to k_r}[\frac{d}{dk}[\Delta_l(k^2)-iC_0^2(\eta)]]}
\end{equation}
Here $E_r=E_0-i\Gamma$/2,  $k_r=\sqrt{2\mu E_r}, k_r=k_0-\it{i}k_i$.

According to the known relations between the NVC ($\tilde{G}_l$), ANC ($C_l$) and the residue $W_l$ we can write
\begin{equation}\label{NVC}
\tilde G_l^2=-\frac{2\pi k_r}{\mu^2}W_l,
\end{equation}
\begin{equation}\label{AbsANC}
C_l=\frac{i^{-l}\mu}{\sqrt{\pi}}\frac{\Gamma(l+1+i\eta_r)}{l!}e^{-\frac{\pi\eta_r}{2}}\tilde G_l
\end{equation}
where $\eta_r=\xi/k_r$.
A simple relation for the ANC derived in Ref. \cite{dolin_akram} for narrow resonances  which we call the Dolinsky-Mukhamedzhanov  (DM) method
\begin{equation}\label{dolin_akram}
\mid C_l^a\mid = \sqrt{\frac{\mu\Gamma}{k_0}}
\end{equation}
 is used to check our calculations.

The uncertainty of the absolute value of the ANC which follows from Eq. (\ref{dolin_akram}) is obtained by deducing a  differential of the right hand side of Eq. (\ref{dolin_akram}) which is the function of the two arguments $E_0$ and $\Gamma$:
\begin{equation}\label{DeltaC}
\Delta C_l =|C_l^a| \left(\frac{\Delta\Gamma}{2\Gamma}+\frac{\Delta E_0}{4 E_0}\right)
\end{equation}
where we put the increment signs instead of the differentials, which should not be confused with the already used signs for the $\Delta$ method and the $\Delta_l(k)$-function. One can see from the last equation that the uncertainty in the width ($\Delta\Gamma$) contributes twice as much compared to that ($\Delta E_0$) in
the resonance energy into the relative ANC uncertainty  $\Delta C_l /|C_l^a|$.

\section{The ANC from the elastic scattering amplitude based on the analytic properties of the $S$-matrix (SMP-method)}\label{S-matrix}

Near an isolated resonance the partial $S$-matrix element can be represented as in Ref. \cite{migdal}
\begin{equation}\label{S-mat}
S_l(k)=e^{2i\nu_l(k)}\frac{(k+k_r)(k-k_r^*)}{(k-k_r)(k+k_r^*)},
\end{equation}
where $k_r=k_0-ik_i$ is the complex wave number of a resonance  ($k_0>k_i>0$, and the symbol (*) means the complex conjugate operation). Here $k_0>k_i$ because we do not consider sub-threshold resonances.  Energy $E_0$ of this resonance and its width $\Gamma$ are
\begin{equation}\label{energy res}
E_0=\frac{k_0^2-k_i^2}{2\mu},\qquad \Gamma=\frac{2k_0k_i}{\mu}.
\end{equation}
The partial scattering nonresonant phase shift $\nu_l(k)$ is a smooth function near the pole of the $S$-matrix element, corresponding to the resonance. The $S$-matrix element defined by Eq. (\ref{S-mat}) fulfills the conditions of analyticity, unitarity and symmetry. It is possible  to recast Eq. (\ref{S-mat}) in the form
\begin{equation}\label{S-phas}
S_l(k)=e^{2i(\nu_l+\delta_r+\delta_a)},
\end{equation}
where $$\delta_r=-\arctan{\frac{k_i}{k-k_0}}$$ represents the resonance phase shift, while $$\delta_a=-\arctan{\frac{k_i}{k+k_0}}$$ is the additional phase shift which  contributes to the whole scattering phase shift. Thus the total phase shift is
\begin{equation}\label{phase}
\delta_l=\nu_l+\delta_r+\delta_a.
\end{equation}

After simplification and replacing exp(2i$\delta_l$) by $S_l(k)$ we get
\begin{equation}\label{fl}
\tilde{f}_l(k)=\frac{S_l(k)-1}{2ik\rho_l(k)},
\end{equation}

This renormalized amplitude $\tilde{f}_l(k)$ can be analytically continued like the partial scattering amplitude, corresponding to the short-range interaction, and has its pole at the point $k_r$ according to Eq. (\ref{S-mat}).  In the vicinity of  pole $k_r$, the partial scattering amplitude (\ref{fl}) can be represented as
\begin{equation}\label{pole}
\tilde{f}_l(k)=\frac{W_l}{k-k_r}+\tilde{f}_{l,nonres}(k),
\end{equation}
where the function $\tilde{f}_{l,nonres}(k)$ is regular at this point.
The simple derivation of the residue $W_l$ leads to the expression
\begin{equation}\label{residue}
W_l=\text{res}\tilde{f}_l=\lim_{k\to k_r}\Bigl[(k-k_r)
\tilde{f}_l(k)\Bigr]= -\frac{k_ie^{i2\nu_l(k_r)}}{k_0\rho_l(k_r)}.
\end{equation}
Using the relationship between the NVC ($\tilde{G}_l$) and ANC ($C_l$) (\ref{AbsANC}), we obtain
\begin{eqnarray}\label{ANC}
C_l=\frac{i^{-l}\mu}{\sqrt{\pi}}\frac{\Gamma(l+1+i\eta_r)}{l!}e^{-\frac{\pi\eta_r}{2}}\tilde G_l\qquad\nonumber\\
=i^{-l}\sqrt{\frac{\mu\Gamma}{k_0}}e^{-\frac{\pi\eta_r}{2}}\frac{\Gamma(l+1+i\eta_r)}{l!}\nonumber\\
\times e^{i\nu_l(k_r)}\sqrt{(1-ik_i/k_0)/\rho_l(k_r)}.\qquad
\end{eqnarray}
The derived equations are valid for both narrow and broad resonances. For narrow resonances, when $\Gamma \ll E_0$ ($k_i \ll k_0$), one can simplify Eq.(\ref{ANC}) for the ANC replacing $k_r$ by $k_0$ and using the equality
\begin{equation}\label{simp}
e^{-\frac{\pi\eta}{2}}\frac{\Gamma(l+1+i\eta)}{l!\sqrt{\rho_l(k_0)}}=e^{i\sigma_l}
\end{equation}
to obtain
\begin{equation}\label{narrow}
C^a_l=\sqrt{\frac{\mu\Gamma}{k_0}}e^{i(\nu_l(k_0)+\sigma_l(k_0)-\pi l/2)},
\end{equation}
which coincides with the result obtained in Ref. \cite{dolin_akram}.

The nonresonant phase shift $\nu_l(k)$ is the analytical function, excluding the origin. In Ref. \cite{Mur83}, the authors present the behavior of $\nu_l(k)$ near the origin as
\begin{equation}\label{nu-l}
\nu_l(k)=-\frac{2\pi}{(l!)^2}k^{2l+1}\eta^{2l+1}a_l e^{-2\pi\eta},
\end{equation}
where $a_l$ is the scattering length for colliding nuclei. We see that $k=0$ is the point of the essential singularity of the scattering phase shift. However, as a function of the momentum $k$, it has normal analytical properties near the point corresponding to the resonance. Therefore we can expand $\nu_l(k)$ to a series
\begin{equation}\label{series}
\nu_l(k)=\sum\limits_{n=0}^{\infty}c_n(k-k_s)^n
\end{equation}
in the vicinity of the pole corresponding to the resonance.
If we wish to determine the value of the phase shift $\nu_l(k)$ by applying Eq. (\ref{series}) at a point in the complex plane close to the centered point $k_s$, then  only the first few terms of the convergent series may be taken into account with reasonable precision.
The expansion coefficients $c_n$ of Eq. (\ref{series}) as well as $k_0$ and $k_i$ are determined by fitting the experimental values of the elastic scattering phase shifts $\delta_l$ given by Eq. (\ref{phase}).

\section{Resonance ANC calculated by $\Delta$ method and its comparison with SMP method results for  low-energy  $^7\rm{\bf {Be}}$, $^8\rm{\bf{Be}}$ and $^ {16}$O  levels}

 As mentioned in the introduction  we apply the $\Delta$ method described above to the $^7$Be,  $^8$Be  and  $^{16}$O resonances using a  model with the configurations $^3$He+$\alpha$,  $\alpha$+$\alpha$ and $\alpha$+$^{12}$C. For the  $\alpha$-$^{12}$C collision  the absolute value of the nuclear part of the ERF is very small compared with those for the Coulomb part (see Ref. \cite{OrlovIrgazievPRC96}) due to the  'large'  product of the colliding particle charges. As explained above, this is due to $\exp(2\pi\eta)$ in the denominator in Eq. (\ref{CandDfunctions}) for $C_0^2$.  Therefore, the ERF approach is not valid for this nucleus.
		
In Ref. \cite{OrlovIrgazievPRC96} we find that the results of the fitting are quite sensitive to the selection of the energy region. For bound states the low-energy area is especially important, while for resonances it is necessary to secure a proper description of $\Delta_l$ in the vicinity of  resonance energy $E_0$.
\begin{figure}[thb]
\parbox{8.0cm}{\includegraphics[width=8.0cm]{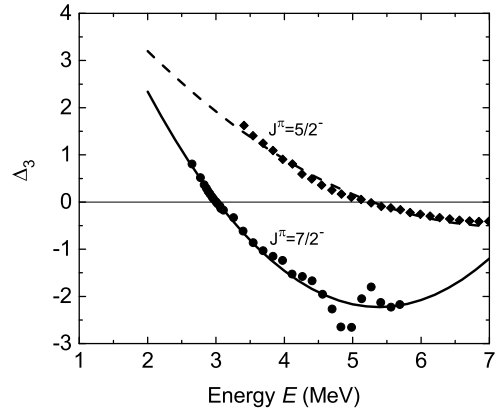}}
\caption{Dependence of the fitted $\Delta_l$ functions [Eq. (\ref{7Be8Be})] vs the center-of-mass energy $E$ of the $^3\rm{He}$-$\alpha$ collision. Solid and dashed lines are for $J^\pi=7/2^-$ and 5/2$^-$, respectively. The experimental data (dots) correspond to the phase shifts taken from Ref. \cite{Spiger}. Results of the extracted resonance energy are $E_0=3.017$ MeV, $\Gamma=177$ keV for $J^\pi=7/2^-$ and $E_0=5.106$ MeV , $\Gamma=1.212$ MeV  for $J^\pi=5/2^-$. \label{fig1}}
\end{figure}	
\begin{figure}[thb]
\parbox{8.0cm}{\includegraphics[width=8.0cm]{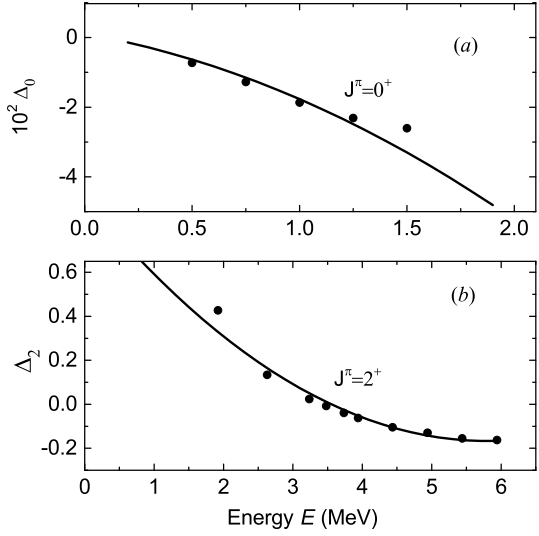}}
\caption{As for Fig. 1 for the $\alpha$-$\alpha$ collision for $J^\pi=0^+$ (\textit{a}), and $ 2^+$  (\textit{b}). The experimental data (dots) correspond to the experimental phase shifts taken from Ref. \cite{ahmed}. The extracted resonance energy and width are: $E_0=0.093$ MeV, $\Gamma=0.0055$ keV for $J^\pi=0^+$ and $E_0=3.096$ MeV, $\Gamma=1.512$ MeV for $J^\pi=2^+$. \label{fig2}}
\end{figure}	
	
We need to satisfy this demand in our calculations while choosing a fitting model.	
In Ref. \cite{OrlovIrgazievPRC96} we use  different methods for the orbital momenta $l$=0, 1, 2 of the bound states.
\begin{figure*}
\parbox{17cm}{\includegraphics[width=17cm]{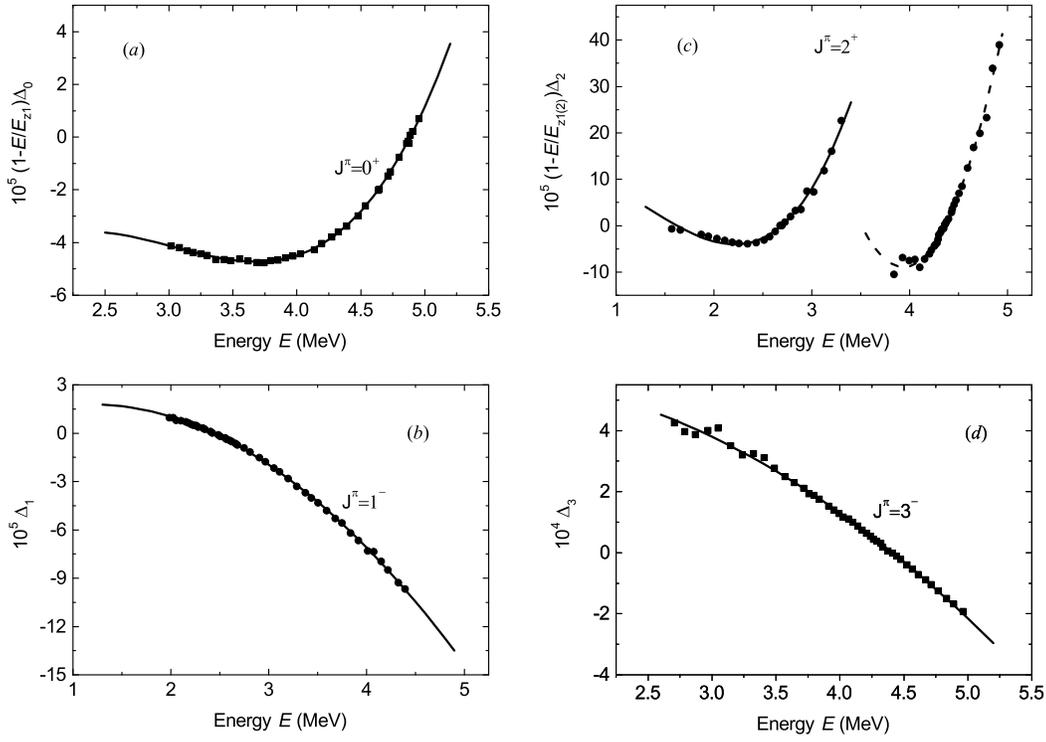}}
\caption{As for Fig. 1 for the $\alpha$-$^{12}\rm{C}$ collision. The fitting methods are given in Eqs.  (\ref{0-0+})--(\ref{3-3-}).  Solid lines are for $J^\pi$=0$^+ (\textit{a}$), $1^- (\textit{b}$), $2^+ (\textit{c}$), left curve, and for $3^- (\textit{d}$). Dashed line is for $J^\pi =2^+ (\textit{c}$), right curve. The state 2$^+$ includes two resonances. The experimental data (dots) correspond to the $R$-matrix fit phase shifts taken from Ref. \cite{tisch}. The extracted resonance energy and width are: $E_0=4.887$ MeV, $\Gamma=3.52$ keV ($0^+$); $E_0=2.364$ MeV,  $ \Gamma=0.333$ MeV ($1^-$); $E_0=2.693$ MeV, $\Gamma=0.597$ keV ($2^+$);  $E_0=4.359$ MeV, $\Gamma=78$ keV ($2^+$);  $E_0=4.228$ MeV, $\Gamma=0.817$ MeV ($3^-$). \label{fig3} }
\end{figure*}	
Here we change the model only for $^{16}$O states to secure a proper agreement of the resonance energy defined from Eq. (\ref{resonance pole condition}) with the experimental values of $E_0$ and $\Gamma$.

The figures for the $\Delta_l$ fit are given for the experimental phase-shift values as well as for $E_0$  and $\Gamma$, written in the corresponding captions for the different nuclear systems and states.

The results in the tables below are given for the different methods and take into account the uncertainties in the resonance energy and in the experimental phase shifts when known.

A simple model of the $\Delta_l$ fit is used for $^7$Be (7/2$^-$,  5/2$^-$) and  $^8$Be  (0$^+$ , 2$^+$):
\begin{equation}\label{7Be8Be}
\Delta_l (E)= a_0+a_1 E + a_2 E^2.
\end{equation}

For $^{16}$O ($^4$He+$^{12}$C) we use the following more complex fitting models:
\begin{enumerate}
\item  $J^{\pi}=0^+$.
There is a narrow resonance in this state (in the $R$ matrix fit in Ref. \cite{tisch} $\Gamma$ = 3 keV). The input phase shift $\delta_0$ is zero or $\pi$ at $E_{z}$=4.8823 MeV.  So the following fitting is used:
\begin{eqnarray}\label{0-0+}
\Delta_0=\qquad\qquad\qquad\qquad\qquad\qquad\qquad\qquad\qquad\nonumber\\
\frac{a_0 + a_1(E-E_{z}) + a_2(E-E_{z})^2 + a_3(E-E_{z})^3}{1-E/E_{z}}.
\end{eqnarray}
\item  $J^{\pi}$=1$^-$.
  Due to the near-threshold bound state  at $E=-\epsilon_1=-$0.045 MeV ($\Delta$=0 at a bound pole \cite{OrlovIrgazievPRC96}) we choose
\begin{equation}\label{1-1-}
\Delta_1=(1+E/\epsilon_1) (a_0 + a_1 E + a_2 E^2).
\end{equation}
\item $J^{\pi}$=2$^+$.
There are two resonances  which are observed in the energy interval 2.5--5.0 MeV.  The bound state pole is situated at $E=-\epsilon_2=-$0.245 MeV. The input phase shift $\delta_2$ is zero or $\pi$ at
$E_{z1}$=2.680 MeV where $\cot\delta_2$ goes to infinity.   Consequently, the fitting model in the region of the lowest resonance may be taken as
\begin{equation}\label{2-2+}
\Delta_2=\frac{(1+E/\epsilon_2)(a_0+a_1 E+a_2 E^2)}{1-E/E_{z1}}.
\end{equation}
The second energy value where $\sin\delta_2$=0 is
$E_{z2}$=3.970 MeV. Therefore,  the fitting model in the region of the second resonance can be taken as
\begin{equation}\label{22-2^+}
\Delta_2=\frac{a_0+a_1(E-E_{z2})+a_2(E-E_{z2})^2}{1-E/E_{z2}},
\end{equation}
where we take $E_{z2}$ as the centered point.
\item $J^{\pi}$=3$^-$. For the fitting we use an expansion with the centered point at $E_{z}$=4.32 MeV in the vicinity of the resonance pole:
\begin{equation}\label{3-3-}
\Delta_3=a_0 + a_1(E-E_{z}) + a_2(E-E_{z})^2.
\end{equation}
\end{enumerate}

In Figs. \ref{fig1}–-\ref{fig3}, we show a comparison between the experimental $\Delta_l$  function values and the fitting curves for the models given in Eqs. (\ref{7Be8Be})--(\ref{3-3-}). The phase-shift experimental data are taken from papers \cite{Spiger} for $^7$Be, \cite{ahmed} for $^8$Be, and \cite{tisch} for the $R$-matrix phase-shift fit for $^{16}$O. There is a fairly good agreement between the fitting curves and the experimental data in the energy intervals considered.
\begin{table}[htb!]
\caption{ $^7$Be$\, \leftrightarrow\alpha+^3$H.   Calculation method, $J^\pi$,  resonance energy  $E_0$ and its width  $\Gamma$, corresponding values  of ANCs $|C_l|$ and  $|C^a_l|$  calculated by Eq. (\ref {dolin_akram}) for narrow resonances. Energy values given in the center-of-mass system. Experimental data \cite{nucldata}: $E_0(5/2^-)=5.143\pm 0.1$ MeV, $\Gamma(5/2^-)=1.2$ MeV; $E_0(7/2^-)=2.983\pm 0.05$ MeV, $\Gamma(7/2^-)=175\pm 7$ keV. }
\begin{ruledtabular}
\begin{tabular}{lccccc}
Method&$J^\pi$ & $E_0$(MeV) & $\Gamma$(keV) & $|C_l|$(fm$^{-1/2})$ & $| C^a_l|$(fm$^{-1/2})$\\
\hline
$\Delta(1)$&$5/2^-$&5.106&1212&0.260&0.277\\
 DM(+)     &               &5.243&1200&0.276&0.274\\
 DM($-$)     &               &5.043&1200&0.274&0.276\\
SMP        &                &4.983&1275&0.264&0.286\\
\hline
$\Delta(1)$&$7/2^-$&3.017&177&0.120&0.121\\
 DM(+) &                   &3.033   &182&0.123&0.122\\
 DM($-$) &                  &2.933   &168&0.118& 0.119\\
SMP&                       &2.987   &182&0.122& 0.123\\
 \end{tabular}
 \end{ruledtabular}
\label{tab1}
\end{table}

\begin{table}[htb!]
\caption{ $^8$Be $\leftrightarrow\alpha+\alpha$.   The definitions of the method, state and the calculated results are the same as in Table I.   Experimental data \cite{nucldata}: $E_0(0^+)=91.84$ keV, $\Gamma(0^+)=5.57\pm 0.25$ eV; $E_0(2^+)=3.122\pm 0.010$ MeV, $\Gamma(2^+)=1.513\pm 0.015$ MeV.}
\begin{ruledtabular}
\begin{tabular}{lccccc}
Method&$J^\pi$ & $E_0$(MeV) & $\Gamma$(keV) & $|C_l|$(fm$^{-1/2})$ & $| C^a_l|$(fm$^{-1/2})$\\
\hline
$\Delta(1)$&$0^+$&0.093&0.0055&0.0016&0.00170\\
ERE(1)     &               &0.0918&0.0058&0.00169&0.00172\\
ERE(2)     &               &0.0918&0.0053&0.00165&0.00165\\
 DM(+)    &               &0.0918&0.0058&0.00172&0.00172\\
 DM($-$)      &               &0.0918&0.0053&0.00165&0.00165\\
SMP        &                &0.0093&0.0056&0.0016&0.00170\\
\hline
$\Delta(1)$&$2^+$&3.096&1512&0.321&0.363\\
$\Delta$(up)&          &2.925&1456&0.329&0.362\\ 
$\Delta$(low)&          &2.899&1669&0.348&0.387\\ 
ERE(1)     &               &2.87&1310&0.348&0.345\\
ERE(2)     &               &2.91&1370&0.361&0.351\\
ERE(3)     &               &3.04&1510&0.387&0.365\\
 DM(+) &                   &3.132   &1528&0.366&0.365\\
 DM($-$) &                  &3.112   &1498&0.362&0.362\\
SMP&                       &3.122   &1513&0.291& 0.362\\
 \end{tabular}
 \end{ruledtabular}
\label{tab2}
\end{table}
\begin{table}[htb!]
\caption{ $^{16}$O $\leftrightarrow\alpha+^{12}$C.   The definitions of the method, state and the calculated results are the same as in Table I.  Experimental data \cite{nucldata}: $E_0(0^+)=4.887 \pm 0.002$ MeV, $\Gamma(0^+)=1.5 \pm 0.5$ keV; $E_0(1^-)=2.423\pm 0.011$ MeV, $\Gamma(1^-)=0.420\pm 0.020$ MeV; $E_0(2^+)=2.683$ MeV$\pm$ 0.5  keV, $\Gamma(2^+)=0.625\pm 0.100$ keV; $E_0(2^+)=4.358$ MeV$\pm$4 keV, $\Gamma(2^+)=71\pm 3$ keV; $E_0(3^-)=4.438$ MeV$\pm$20 keV, $\Gamma(3^-)=0.800\pm 0.1$ MeV. }
\begin{ruledtabular}
\begin{tabular}{lccccc}
Method&$J^\pi$ & $E_0$(MeV) & $\Gamma$(keV) & $|C_l|$(fm$^{-1/2})$ & $| C^a_l|$(fm$^{-1/2})$\\
\hline
$\Delta(1)$&$0^+$&4.887&3.52&0.0174&0.0174\\
DM(+)     &               &4.889&2.0&0.0132&0.0131\\
DM($-$)     &               &4.885&1.0&0.0094&0.0093\\
SMP        &                &4.887&3.0&0.0160&0.0160\\ 
\hline
$\Delta(1)$&$1^-$&2.364&333&0.179&0.185\\
$\Delta$(up)&          &2.213&319&0.178&0.202\\ 
$\Delta$(low)&          &2.327&323&0.177&0.200\\ 
SMP&                       &2.364  &356&0.185& 0.209\\
\hline
$\Delta(1)$&$2^+$  &2.693&0.597&0.0083&0.0085\\
DM(+)           &           &2.683&0.725&0.0092&0.0092\\
DM($-$)          &           &2.682&0.525&0.0078&0.0078\\
SMP               &           &2.364&0.760&0.0094&0.0094\\
\hline
$\Delta(1)$&$2^+$  &4.359&78&0.0835&0.0842\\
$\Delta$(up)&           &4.380&80.35&0.0846&0.0853\\ 
$\Delta$(low)&          &4.386&73.84&0.0810&0.0817\\ 
SMP&                         &4.350  &79.1&0.0840&0.0847\\
\hline
$\Delta(1)$&$3^-$  &4.228&817&0.236&0.274\\
$\Delta$(up)&           &4.266&809&0.240&0.272\\ 
$\Delta$(low)&          &4.257&825&0.238&0.275\\ 
SMP&                         &4.350  &79.1&0.230&0.273\\
 \end{tabular}
 \end{ruledtabular}
\label{tab3}
\end{table}
In the Tables \ref{tab1}--\ref{tab3} all the experimental data for the energies and widths of the resonances are taken from Ref. \cite{nucldata}. The left column 'Method' in all  tables includes the following designations.
$\Delta(1)$ means the calculation by the $\Delta$ method, using  the models given above with the experimental phase-shift values.
$\Delta$(up) and $\Delta$(low) mean the calculation by the $\Delta$ method, using  the upper and lower experimental phase-shift values taken from Ref. \cite{ahmed} (the $\alpha$--$\alpha$ scattering phase shifts) and Ref. \cite{plaga87} (the $\alpha$--$^{12}$C scattering phase shifts).

The rows DM(+) and DM(-) show the results of the calculations by the Eq. (\ref{dolin_akram}) for narrow resonances, where '+' and '$-$' are related to the  maximal and minimal values of the experimental energy and width of the  resonance in accordance with the uncertainty defined by the Eq. (\ref{DeltaC}), respectively.
In Tables I and II, the row denoted by SMP shows our new calculations for  $^7$Be and $^8$Be using the SMP-method. In Table III, all the SMP-method results are taken from our paper \cite{irgaz-14}.
In Table II, for the ground 0$^+$ state the results noted ERE(1) and ERE(2) are calculated by the conventional method using the $G_l^2$ results found in  Ref. \cite{PHAN77} for differing resonant energies. The corresponding results for the 2$^+$ state are noted by the ERE(1, 2, 3). These ERF-method results also show the effects of the resonant energy and width variations. One can compare these ERF results with those using other methods.

Analysis of the calculation results in all the tables leads to the following conclusions.
\begin{enumerate}
\item For narrow resonances, different accepted methods lead to  good agreement between the $|C_l|$ and $| C^a_l|$. Such agreement can be considered as a criterion of  a resonance narrowness. This is slightly better for $\Delta(1)$ compared with the  SMP-method results.
\item For wider resonances, as a rule there is an inequality $|C_l|<|C^a_l|$, but the differences are not very big for the nuclear systems considered. We include the results of $|C_l^a|$ for broad resonances to find out how large the differences are between $|C_l^a|$ and $|C_l|$.
\item There is very good agreement between $|C_l|$ and $|C^a_l|$ for both DM(+) and DM($-$) in Tables I and II with one exception in Table III for $J^\pi=0^+$. This is due to the high sensitivity of $|C_l|$ to the $\Gamma$ value. Significantly bigger variations of the resonant width are visible in this case, which lead to the variation $|C_0|$=0.0094--0.0174 fm$^{-1/2}$ although the difference between
 $|C_0|$ for $\Delta(1)$-model (0.0174 fm$^{-1/2}$) and SMP-method (0.0160 fm$^{-1/2}$) is rather small. In Table I for $J^\pi=5/2^-$
$E_0$ changes while $\Gamma$ is mostly stable. This lead to the variation
$|C_2|$=0.260--0.277  fm$^{-1/2}$. For the narrow resonance in the $7/2^-$ state which is most pronounced experimentally, the variation is smaller
$|C_3|$=0.118--0.123  fm$^{-1/2}$ fm$^{-1/2}$. We note that in Ref. [26] the resonance energy $E_0(2^+) = 3.03 \pm 0.01$ MeV, $\Gamma = 1.49 \pm 0.02$ MeV.  These $E_0(2^+)$ and $\Gamma$ are slightly smaller compared with the experimental data [23] as are some values given in Table \ref{tab2}.
\item Table \ref{tab2} shows for the $\alpha$-$\alpha$ system good agreement ($|C_0|$=0.0016--0.0017 fm$^{-1/2}$)
 for all the models, including the ERF for the  $J^\pi=0 ^+$  state, which is especially important in astrophysics. For the 2$^+$ state, the experimental uncertainties in the resonant energy and the phase shift lead to the variation $|C_2|$=0.32--0.35  fm$^{-1/2}$.
\item Table III for the $\alpha + ^{12}$C  system is the  most important in the present paper because the ERF-method is invalid. It contains much more information than the other tables, including the effects of uncertainties of the phase shifts for all the states considered except the  $0^+$ and  first 2$^+$ resonance states.
In  the second state 2$^+$ where  we find a stability of  $E_0$ but also an essential $\Gamma$ variation which  significantly affects the value of the $|C_l|$.  Nevertheless, there is quite good agreement between the results for  the $\Delta(1)$ and the SMP methods.
\end{enumerate}

 \section{Conclusion}\label{conclusion}
In the present paper we apply the $\Delta$ method to resonances.
We use the original form (\ref{fl2}) of the renormalized scattering amplitude where there is no expression for the ERF.
We emphasize that the renormalized scattering amplitude does not include the Coulomb term $h(\eta)$ (\ref{Coulomb_h}) which is part of the ERF in Eq. (\ref{CoulombKl}).
The function $h(\eta)$ forms a background for the nuclear term and obviously leads to a wrong ANC for 'large' charges of colliding nuclei. The $^{16}$O states with the configuration $\alpha + ^{12}$C is an example.
We show that the $\Delta$ or SMP methods should be applied when the calculations using the ERF fitting are invalid. In Table III, the calculation results denoted by SMP are taken from Ref. \cite{irgaz-14}.

We include some uncertainties of the experimental
data in our ANC calculations. The formula for
narrow resonances is used to derive a simple
expression for the increment $\Delta C_l$ related to both
the uncertainties of the resonance energy ($\Delta E_0$)
and width ($\Delta\Gamma$). Some experimental uncertainties
of the phase-shift data are also included and their effects in ANCs are analyzed.
The system $\alpha +^{12}$C is studied in more detail using
the $\Delta$ method as the conventional ERF-method is not valid for this system or for those
with larger charge product. We also study the
lighter systems $^3$He$^4$He and $\alpha\alpha$. The ground state
of  $^8$Be is especially important in astrophysics
for the creation of the organic elements and
life itself on Earth \cite{Anthropic}. While considering the different pole conditions for bound and resonance state, we stress the role of the square root cut on the complex energy plane of the partial scattering amplitude. In addition, the renormalized amplitude is real on the imaginary momentum axis. This is due to the similarity of its analytical properties to those of the amplitude for a short-range potential (see, for example, Ref. \cite{Taylor}).

The reasonable agreement between the resonant energies and the ANC results obtained by both the $\Delta$  and SMP methods, as well as that between the ANC results and the ANCs calculated for narrow resonances mean that these results are credible for the nuclear systems considered within the found limits of the variations.
They can be used in nuclear astrophysics and in the nuclear
reactions theory based on Feynman diagrams.

\section*{ACKNOWLEDGEMENTS}
This work was partially supported by the Russian Foundation for Basic Research Grant No. 16-02-00049.
The authors are grateful to H.M. Jones for editing the English.

\end{document}